\documentstyle[12pt,psfig]{article}
\advance\textheight by\topskip
\newcommand{\be}{\begin{eqnarray}}
\newcommand{\ee}{\end{eqnarray}}
\newcommand{\bi}{\bibitem}
\newcommand{\rar}{\rightarrow}

\newcommand{\nuh}{\nu_h}
\newcommand{\mnh}{m_{\nu_h}}

\input{epsf}

\begin{document}


\title{ 
DARK MATTER IN THE UNIVERSE 
}

\author{     A.D. Dolgov
\and{      }
\and{\it  TAC, Juliane Maries Vej 30, DK-2100, Copenhagen, Denmark       }
\and{\it Also: ITEP, Bol. Cheremushkinskaya 25, Moscow 113259, Russia }   
\and{\it                   E-mail: dolgov@tac.dk                      }
}
\date{}
\maketitle

\begin{abstract}

Cosmological arguments proving that the universe is dominated by invisible
non-baryonic matter are reviewed. Possible physical candidates for dark
matter particles are discussed. A particular attention is paid to 
non-compensated remnants of vacuum energy, to the question of stability of
super-heavy relics, cosmological mass bounds for very heavy neutral lepton, 
and some other more exotic possibilities.

\end{abstract}

\section {Introduction \label{intr}}

Probably one of the most important discoveries of this Century was
the discovery that the universe consists mostly of an unknown 
form of matter. This matter neither emit nor absorb light and got the name
dark (or better to say, invisible) matter. It is observed only indirectly
through its gravitational action and, though there are plenty of theoretical
hypotheses, the nature of dark matter remains mysterious. First hints 
on existence of dark matter were found more than half of a century 
ago~\cite{oort32,zwicky33}. Velocity dispersion of astronomical objects was 
larger than one would expect from observation of luminous 
matter. The fact that there is more mass than light in the
universe, got a strong support only 40 years later. It was initiated by two 
groups~\cite{einasto74,ostriker74} and stimulated a burst of activity in the 
field. Now there are a large amount of accumulated astronomical data that
unambiguously prove that the universe is dominated by an invisible matter or
to be more precise there is much more gravity in the universe than all 
the visible matter could provide.

\begin{figure}[ht]
\centerline{\psfig{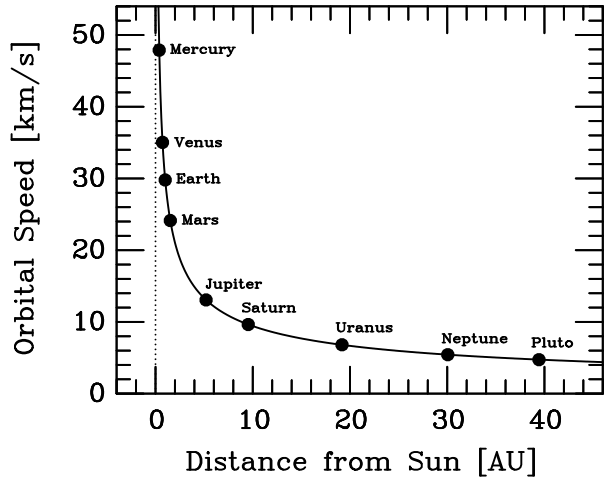}}
\caption{Rotation curve of the solar system which falls off as
$1/\protect\sqrt{r}$ in accordance with Kepler's law.
The astronomical unit (AU) is the Earth-Sun distance of
$1.50\times10^{13}\,\rm cm$.
\label{fig:kepler}}
\end{figure}

Very strong arguments in favor of invisible cosmic matter follow from the
so called galactic rotational curves, i.e. from the observed dependence of 
velocities of gravitationally bound bodies on the distance from the visible
center. A very well known example of rotational curves that have led to the
seminal discovery of the Newton gravitational law is the distribution of
velocities of planets in the Solar system (see fig.~\ref{fig:kepler}, taken 
from ref.~\cite{raffelt97}).

\begin{figure}
\par
\centerline{\vbox{
\psfig{figure=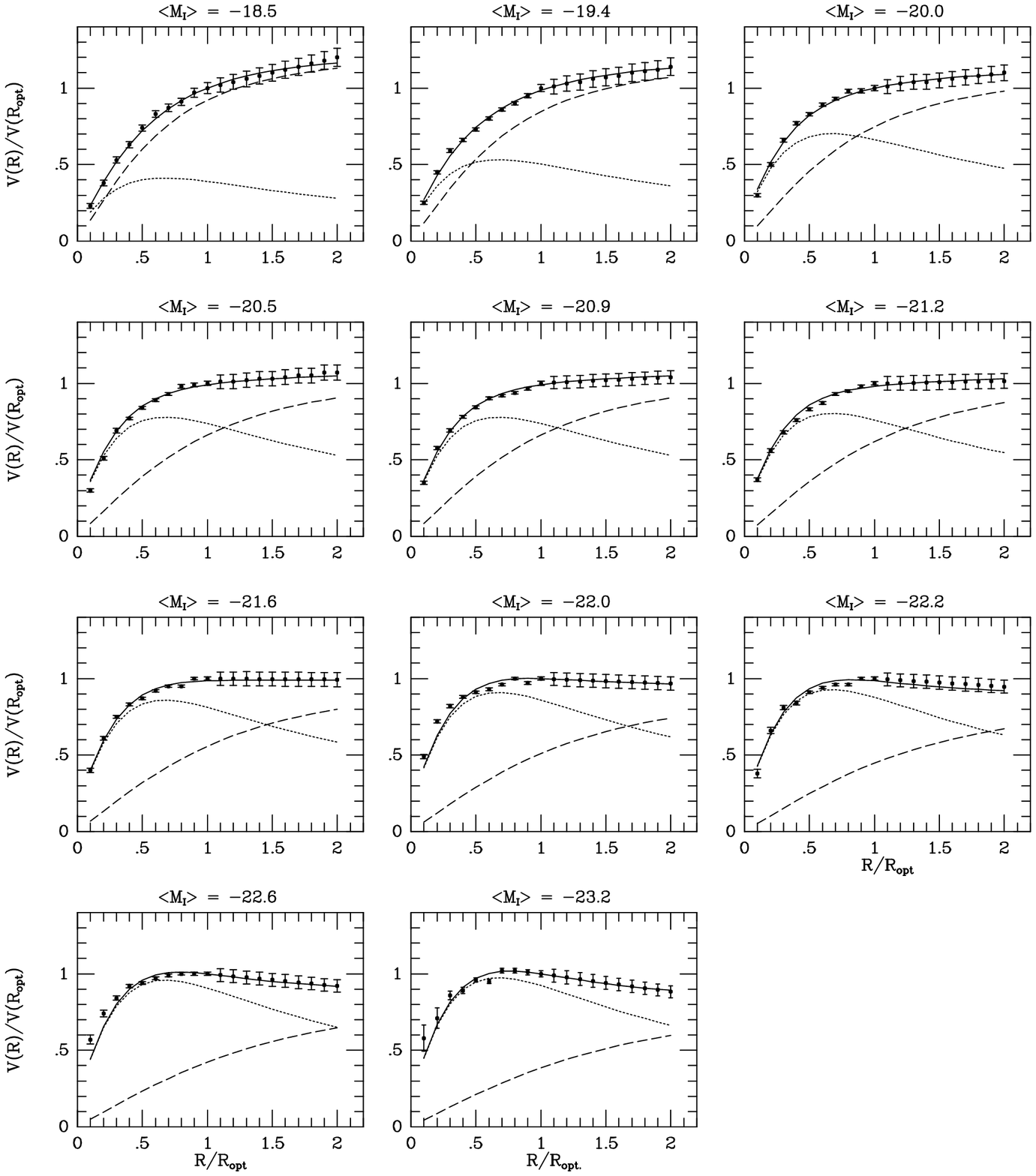,height=8.8cm,width=5.6cm} }}
\par
\caption { Coadded rotation curves (filled circles with error bars)
reproduced by universal rotational curve (solid line) Also shown the 
separate dark/luminous contributions (dotted line: disk;  dashed line: halo.)}
\label{fig:flat}
\end{figure}

On the basis of this data it was concluded  that gravitational forces drops
down with distance as $F\sim 1/r^2$ and correspondingly, by the virial theorem,
$v^2 (r) \sim G_N M(r)/r$, so that $v\sim 1/\sqrt{r}$ for point-like central
mass; here $M(r)$ is the mass of gravitating matter inside the radius $r$.
However measurements of rotational velocities of gas around galaxies
produce a very different picture, $v(r)$ does not go down to zero with
an increasing distance from the luminous center but tends to a 
constant value, see fig.~\ref{fig:flat}~\cite{salucci97}. 
At the present
day more than 1000 galactic rotational curves are measured (see 
e.g.~\cite{persic95}) and they show a similar behavior. It is quite
a striking fact that rotational curves are very accurately flat at large
distances, $v \rar const$.
If such curves were observed at Kepler-Newton's time one might conclude that
the gravitational force did not obey the famous inverse square law but
something quite different, $F\sim 1/r$, with the potential $U\sim \ln r$. 
However it is very difficult, if possible at all, to modify beautiful general 
relativity at large distances in such a way that it would give $1/r$-forces.
A normal interpretation of flat rotational curves is that there is an invisible
matter around galaxies with mass density decreasing as 
\be
\rho\sim 1/r^2
\label{rhoofr}
\ee
and correspondingly $M(r) \sim r$. Such mass distribution could be in a
self-gravitating isothermal gas sphere. However, if the dark matter
particles do not possess a sufficiently strong self-interaction it is not
clear how they would acquire thermal equilibrium.

It is not yet established how far the law (\ref{rhoofr})
remains valid. If it is true up to neighboring galaxies, the average mass
density of this invisible matter would be rather close to the critical one
\be
\rho_c = {3 H^2 \over 8\pi G_N } 
\approx 1.86\cdot 10^{-29} h_{100}\,\, {\rm g/cm}^3
\label{rhoc}
\ee
where $h_{100} = H/ 100 {\rm km/s/Mpc}$ is dimensionless Hubble constant;
by the most recent data~\cite{hubble} $h_{100}\approx 0.7$ with the
error bars of about 10-15\%; for a review see ref.~\cite{freeman99}.

The contributions of different forms of matter to the cosmological 
mass/energy density according to the present day data is the 
following. The visible luminous matter contributes very little to total
density~\cite{persic92}:
\be
\Omega_{lum}= \rho_{lum} / \rho_c \leq 0.003\, h_{100}^{-1}
\label{omegalum}
\ee 
There could be much more non-luminous baryons in the forms of faint stars, 
gas, etc (see below sec.~\ref{baryon}) but the standard theory of
primordial nucleosynthesis does not allow too high mass fraction of
baryonic matter. It is probably a proper time and place to mention that
George Gamow~\cite{gamow46} made a pioneering contribution to big bang 
nucleosynthesis. Abundances of light elements  
are sensitive to total number fraction of cosmic baryons, more
precisely abundances of light elements depend upon the ratio of number
densities of baryons to photons, 
$\eta_{10}= 10^{10}n_b/n_\gamma$. Comparing theoretical
predictions with observations one can deduce the value of this ratio at
nucleosynthesis. The result crucially depends upon the observed abundance
of deuterium since the latter is especially sensitive to $\eta$. There are
two conflicting pieces of data: high and low deuterium, see discussion and
references in the review ~\cite{olive99}. For low $^2H$ regions the limits
presented in ref.~\cite{olive99} are: 
\be
\Omega h_{100}^2 = 0.015 - 0.023 \,\, {\rm and}\,\, \eta_{10} = 4.2-6.3 
\label{lowd}
\ee
while for high $^2H$:
\be
\Omega h_{100}^2 = 0.004 - 0.01\,\, {\rm and}\,\, \eta_{10} = 1.2-2.8
\label{hid}
\ee 
Most probably one or other of the above is incorrect and the predominant 
attitude is in favor of low deuterium. However, it would be extremely 
interesting if both are true, so that the abundance of primordial deuterium 
is different 
in different regions of the universe. A possible explanation of this 
phenomenon is a large and spatially varying neutrino degeneracy that predicts
a large mass fraction of primordial helium, more than 50\%, comparing to
$\sim 25\%$ in normal deuterium regions (that were called "low" above),
and quite low helium, $\leq 12\%$, 
in the anomalously low deuterium regions~\cite{dolgov98}.  

Anyhow, independently of these subtleties, big bang nucleosynthesis strongly
indicates that the mass fraction of normal baryonic matter in the universe 
is quite small (see also the discussion below in sec.~\ref{baryon}).
On the other hand, the amount of gravitating matter, found by different 
dynamical methods (for a review see~\cite{freedman99}), 
gives $\Omega_m \sim 0.3$.
These methods are sensitive to clustered matter and do not feel uniformly
distributed energy/mass density. Theoretical prediction based on inflationary
model is $\Omega_{tot} = 1 \pm 10^{-4}$. This number may be compatible with the
above quoted value for $\Omega_m$ only if the rest of matter is uniformly 
distributed. The recent indications to a non-zero cosmological 
constant~\cite{perl98} with
\be
\Omega_{vac} \approx 0.7
\label{omegavac}
\ee
permit to fill the gap between 0.3 and 1.
It is possibly too early to make a definite conclusion, since the 
result is very important and all possible checks should be done. Moreover
the SN-data that led to the conclusion of non-zero $\Lambda$ might be
subject to a serious criticism~\cite{riess99}. Still the  combined different
astronomical data quite strongly suggest that cosmological constant is indeed
non-zero. 

The attitude to a possibly non-vanishing cosmological constant from different 
prominent cosmologists and astrophysicists were and is quite diverse. 
For example Einstein, who "invented" cosmological constant and introduced 
it into general relativity, considered it as the biggest blunder of his life.
The attitude of Gamow was similar, he wrote in his autobiography 
book~\cite{gamow70}: "$\lambda$ again rises its nasty head"  
On the other hand, Lemaitre and Eddington considered $\Lambda$ very 
favorably. Moreover, 
a non-zero $\Lambda$ (or what is the same, vacuum energy) should be 
quite naturally non-zero from a particle physicist's point of view, though
any theoretical estimate by far exceeds astronomical upper limits (see
discussion in sec.~\ref{lambda}).

To conclude, it seems very probable that the normal baryonic matter contributes
only a minor fraction to the total mass/energy of the universe and we will 
discuss below possible forms of this yet unknown but dominant part of our
world. It is not excluded that there is not a single form of dark matter. The 
data request several different ones and if it is indeed the case the mystery
becomes even deeper. In particular, one has to understand the so called
cosmic conspiracy: why different forms of dark matter give comparable 
contributions to $\Omega$, while they naturally would differ by many orders
of magnitude.

\section { Baryonic dark matter. \label{baryon}}

Since an idea that there is a cosmic ocean of an absolutely unknown matter
is quite drastic, one is inclined to look for less revolutionary explanations
of the data. The first natural question is if all the dark matter, 
possibly excluding  vacuum
energy, could be the normal baryonic staff somehow hidden from observation.
The relevant discussion of the cosmic baryon budget can be found in 
ref.~\cite{fukugita97}

As we have already mentioned in the Introduction a very strong upper limit
on the total amount of baryons in the universe follows from the Big Bang
Nucleosynthesis. However this limit would be invalid if for example
electronic neutrinos are strongly degenerate~\cite{olive91,kang92}. 
A charge asymmetry in electronic neutrinos corresponding to dimensionless
chemical potential $\mu_{\nu_e}/T \sim 1$ could significantly loosen the
bound on baryonic mass density and make it close to the necessary 
$0.3 \rho_c$. 

However there are some other data that make it very difficult to have
baryon dominated universe. Strong arguments against this possibility
come from the theory of large scale structure formation. In the case of 
adiabatic perturbations that are characterized by approximate equality of 
density and temperature fluctuations, $\delta \rho/\rho \sim \delta T /T$,
there is too little time for cosmic structures to evolve. Indeed the
perturbations in the baryonic matter could rise only after hydrogen 
recombination that took place rather late at redshift $z\approx 10^3$.
After that the perturbations might rise only as the scale factor so to the 
present time they at most could  be amplified by this factor of $10^3$. 
However, it is well known that the fluctuations of the CMB (cosmic 
microwave background) temperature 
are quite small, $\delta T /T < {\rm a \,\, few} \times 10^{-5}$. Hence
even today the density fluctuations should be quite small in contrast to the
observed developed structures with $\delta\rho/\rho \gg 1$.

For isocurvature perturbations the fluctuations of CMB temperature
are much smaller than density perturbations, $\delta T /T \ll \delta \rho/\rho$,
and this permits to avoid the above objection. However if it were the
case, the spectrum of angular fluctuations of CMB would be quite
different from the observed one. In particular, the first acoustic peak 
would be near $l=400$, while the data strongly indicates that this peak
is close to $l=200$ in agreement with adiabatic theory (for a recent review 
and the list of references see e.g. ref.~\cite{rocha99}). This argument
can be avoided if the shift of the acoustic peak to higher $l$ is 
compensated by the curvature effects (I thank J. Silk for indication to this
point).

Another weighty argument against baryonic universe is that it is practically 
impossible to conceal 90\% of baryons. Baryonic matter strongly interacts 
with light and even if the baryons are nonluminous themselves, they 
would strongly absorb light. So baryonic matter should be observed either 
in emission or absorption lines.
There is not much space for baryons to escape detection:
\begin{enumerate}
\item{}
{\it Cold gas or dust} do not emit light but can be observed by absorption 
lines (Gunn-Peterson test).
\item{}
{\it Hot gas} is seen by X-rays if it is clumped, if it is diffuse it would
distort CMB spectrum. 
\item{}
{\it Neutron stars or "normal" black holes"} that were produced as a result
of stellar evolution, would contaminate interstellar medium by "metals"
(elements that are heavier than $^4 He$).
\item{}
{\it Dust} is seen in infrared
\end{enumerate}
According to ref.~\cite{fukugita97} the total baryon budget is in the range:
\be
0.007\leq \Omega_B \leq 0.041
\label{omegab}
\ee
with the best guess $\Omega_B = 0.021$ (for $h_{100}= 0.7$).

A special search was performed for the so called  {\it MACHO's} (massive 
astrophysical compact halo objects). They may include
brown dwarfs, low luminosity stars, primordial black holes. Such objects are
not directly visible and they were looked for through gravitational 
micro-lensing~\cite{pac92}. The search was pioneered by MACHO~\cite{macho} 
and EROS~\cite{eros} collaborations and at the present time about a hundred 
of such objects were found in the Galaxy and in the nearby halo. 
According to the EROS results the mass density of the micro-lenses with 
the masses in the interval $(5\cdot 10^{-8} -10^{-2})\,M_{\odot}$ is less
than $0.2\rho_{Halo}$. The MACHO observations permit to make the conclusion
that the masses of micro-lensing objects lies in the interval 
$(0.1-1.0)\,M_{\odot}$ at 90\% CL. The mean value of the mass is about
$ 0.5 M_{\odot}$. 

Instead of approaching to the resolution of the problem of dark matter, 
these observations made things even more mysterious and more interesting.
A large mass of MACHO's suggests that they could be the remnants of the
usual stars (white dwarfs?). However it is difficult to explain their 
relatively large number density and distribution. They could be primordial
black holes but in this case they are not necessarily baryonic. An intriguing
possibility is that they are the so called mirror or shadow stars, i.e
they are formed from a new form of matter that is related to ours only
gravitationally and possibly by a new very weak interaction (see 
sec.~\ref{shadow}).

Anyhow, baryons seem to contribute only a minor fraction to the total mass
of the universe and some new form of matter should exist. There is no 
shortage of possible candidates but it remains unknown what one (or maybe ones)
is (are) the real dominating entity.

\section{Non-baryonic (exotic?) dark matter; what is it? \label{nonbar}}

For an astronomer the classification of dark matter from the point of view 
of large scale structure formation is especially relevant. Independently 
of its physical nature cosmological dark matter can be of the following three
types:
\begin{enumerate}
\item{}
Hot dark matter (HDM). For this form of dark matter the structure can be
originally formed only at very large scales, much larger than galactic
size, $l_{str} \gg l_{gal}$.
\item{}
Cold dark matter (CDM). It is an opposite limiting case for which the 
structure is formed at the low scale, $l_{str} \ll l_{gal}$.
\item{}
Warm dark matter (WDM). This is an intermediate case when the characteristic
scale of the structures is of the order of galactic size,
$l_{str} \sim l_{gal}$. 
\end{enumerate}
Somewhat separately there stands $\Lambda$-term or, what is the same,
vacuum energy.
There are some rather strong indications that for a good description of the 
observed large scale structure several different forms
of dark matter, including $\Lambda$-term, may be necessary.

Another astronomically important feature of dark matter is its dissipation
properties. If dark matter easily loose energy, the structure formation could 
proceed faster. In the opposite case the cooling of dark matter would 
be less efficient and the structures on small scales would not be formed.
So from this point of view there could be two forms of dark matter,
{\it dissipationless} and/or {\it dissipative}. The dominant part of physical
candidates for dark matter particles are weakly interacting and thus
dissipationless. However there are some, possibly more exotic, models
supplying strongly interacting dark matter particles that could easily loose 
energy.

There are quite many physically possible, and sometimes even natural,
candidates for dark matter particles. An abridged list of them in the
order indicating the author's preference is the following:
\begin{enumerate}
\item{}
Massive neutrinos.
\item{}
Non-compensated remnant of vacuum energy.
\item{}
New not yet discovered, but theoretically predicted, elementary particles:
lightest supersymmetric particle, axion, majoron, unstable but long-lived 
particles, super-heavy relics, etc. It is even possible to construct models
in which the same kind particles would contribute e.g. both to hot and warm 
dark matter.
\item{}
New shadow or mirror world.
\item{}
Primordial black holes.
\item{}
Topological defects (topological solitons).
\item{}
Non-topological solitons.
\item{}
Neither of the above.
\end{enumerate}
It is quite possible that the last entry at the bottom of the list may happen
after all to become the first.

\section{Vacuum energy \label{lambda}}

The problem of vacuum energy is possibly the most striking in the contemporary
physics. Any reasonable theoretical estimate disagree with the astronomical
upper limits on $\rho_{vac}$ by 50-100 orders of magnitude (for a review
see refs.~\cite{weinberg89,dolgov97a}). In fact there
are practically experimentally proven contributions into vacuum energy
from the known in quantum chromodynamics (QCD) vacuum condensates of 
quarks and gluons. The existence of these condensates is necessary for correct
description of hadron properties. In this sense the existence
of these condensates is an {\it experimental
fact}. So we have a fantastic situation: there are well established
contributions into vacuum energy that are larger than the permitted value
by the factor $ 10^{47}$. It may only mean that there is some extremely
accurate mechanism that compensates this huge amount practically down to zero.
Here "zero" is in the scale of elementary particle physics; on astronomical 
scale the remaining vacuum energy may be quite significant. This compensation
should be achieved by something that is not directly related to quarks and 
gluons because all the light fields possessing QCD interactions are known,
while heavy fields cannot make a compensation with the desired accuracy.

It is tempting to assume that the curvature of space-time created by vacuum 
energy would generate a vacuum condensate of a new massless (or extremely light)
field $\Phi$ and the energy of the condensate would cancel down the original 
vacuum energy in accordance with the famous Le Chatelier's principle. It is 
closely analogous to the axionic mechanism of natural CP-conservation in QCD. 
Generic features that one should expect from such compensating (or adjustment)
mechanism are quite interesting. First, the compensation is never complete,
the amount of non-compensated vacuum energy is always parametrically of the
order of critical energy:
\be 
\Delta \rho_{vac} = \rho_{vac}^{in} -\rho_\Phi \sim ({m_{Pl}^2 / t^2}),
\label{deltarhovac}
\ee
but the coefficient of proportionality may be different at different stages of 
the evolution of the universe (e.g. at MD- and RD-stages). Another unusual
feature is that the equation of state of the dark matter corresponding to
$\Delta \rho_{vac}$ may be very much different from the standard ones, 
$p=\rho /3$ at RD-stage or $p=0$ at MD-stage. 

So hopefully such compensating mechanism may be able not only to cut
the "nasty head of $\lambda$" (using Gamow words) but also to extinguish it
almost down to nothing with only a small tail remaining. In fact, it is 
exactly this small tail that induced such a strong negative reaction from 
Gamow, because it could be 100\% cosmologically relevant. This demonstrates
two sides of the cosmological constant problem. Astronomers put the question
if it is cosmologically important, i.e. if $\rho_{vac}$ is not negligible
in comparison with $\rho_c$. If the answer is affirmative, then another 
puzzling problem appears: why vacuum energy, that remains constant in the
course of the cosmological expansion, is close today to $\rho_c$ which evolves 
as $1/t^2$? Particle physicists are more puzzled by the question why vacuum 
energy does not exceed $\rho_c$ by almost an infinite amount. However if this 
is somehow arranged, then the natural value should be precisely zero. So
astronomical indications that $\rho_{vac}$ may be non-vanishing are of
prime importance for all members of astro-particle community.

The compensation mechanism would successfully address both issues: it permits 
to compensate $\rho_{vac}$ to cosmologically acceptable value and gives
a non-compensated remnant of the order of $\rho_c$ at any period of the
history of the universe. 
However all that are predictions of a non-existing theory. Original 
compensating mechanism~\cite{dolgov82} is based on a massless scalar field 
with the Lagrangian:
\be
{\cal L}_0 = \left( \partial \Phi \right)^2 +\xi R \Phi^2
\label{l0}
\ee 
where $R$ is the curvature scalar. For a certain choice of the sign of the 
constant $\xi$ the field $\Phi$ becomes unstable in De Sitter background
(the term $\xi R^2$ behaves as a negative mass squared) and a vacuum 
condensate of $\Phi$ would  evolve. The back-reaction of this condensate 
on the expansion results in a change from the exponential De Sitter regime
to a more slow Friedman one, $a(t)\sim t^\alpha$. So far so good, but this
change of the regime was not achieved by the compensation of the vacuum 
energy. In fact the energy-momentum tensor of $\Phi$ does not have the vacuum 
form, it is not proportional to the metric tensor $g_{\mu\nu}$. The slowing
of the expansion is achieved by the decrease  of the gravitational 
coupling constant with time, $G_N \sim 1/t^2$. 

Other possible candidates on the role of the compensating field could be
fields with higher spins, vector or tensor ones~\cite{dolgov97b}.
More promising seems to be symmetric tensor field $\Phi_{\mu\nu}$. Even the 
simplest possible Lagrangian:
\be
{\cal L}_2 =  \Phi_{\mu\nu;\alpha} \Phi^{\mu\nu;\alpha}
\label{l2}
\ee
gives rise to unstable solution of equations of motion and to development
of vacuum condensate that compensates vacuum energy. In contrast to the
energy-momentum tensor of the considered above scalar field, the 
energy-momentum tensor of $\Phi_{\mu\nu}$ is of vacuum form, i.e. proportional
to $g_{\mu\nu}$.
Such a theory possesses a symmetry with respect to transformation 
$\Phi_{\mu\nu} \rar \Phi_{\mu\nu} + C\, g_{\mu\nu}$. This symmetry prevents 
from quantum generation of mass of $\Phi_{\mu\nu}$ and may be helpful in
some other respects. Still in the simplest versions of the model the 
gravitational coupling constant evolves with time in the same way as in the
scalar field case~\cite{rubakov99}. Presumably it is related to the breaking
of Lorents invariance by the condensate. The model permits a generalization
such that the vacuum field $\Phi_{\mu\nu}$ is proportional to the metric
tensor, $g_{\mu\nu}$, so that the condensate is Lorents invariant. However
in any case the cosmology is far from being realistic. Thus, though the 
compensation mechanism shows some nice features, no workable
model giving realistic cosmology is found at the present day.

Stimulated by the indications that the universe may expand with acceleration,
i.e. that $\rho_{vac} > 0$, a new {\it constant} parameter $w$, was introduced 
into the standard set of cosmological parameters~\cite{caldwell98}. 
This parameter characterizes 
the equation of state of the cosmological matter:
\be
p = w \rho
\label{pw}
\ee
In the standard cosmology it is assumed that the universe is now dominated by
non-relativistic matter, so that $w=0$. At an earlier stage relativistic 
matter was dominating and $w=1/3$. In the case of dominance of vacuum energy
$w=-1$. Two more examples giving a negative $w$ are the system of 
non-interacting cosmic strings with $w=-1/3$ and also non-interacting domain 
walls with $w=-2/3$.
Since the source of gravity in General Relativity (in isotropic case)
is $(\rho + 3 p)$, the universe would expand with acceleration (anti-gravity)
if $w<-1/3$. 

In particular a model with a massless or extremely light scalar field was 
discussed that could give a negative $w$. Such field received the name
"quintessence". For a homogeneous scalar field $\phi (t)$ with a 
self-interaction potential $U (\phi)$ the parameter $w$ is given by:
\be
w= -{ 2 U(\phi) - \dot \phi^2 \over  2 U(\phi) + \dot \phi^2}
\label{wforphi}
\ee
If the potential energy is larger than the kinetic one, $w$ would be negative.
However in this model $w$ may be considered as a constant only approximately. 
A fundamental theory that requests an existence of such a field is missing
so such model can be considered as a poor man phenomenology describing an 
accelerated expansion, more general than just that given by vacuum energy. 
A {\it raison d'\^etre} for such a field could be the adjustment mechanism 
discussed above, that predicts an existence of non-compensated vacuum energy
with an unusual equation of state. Simultaneously, as mentioned above, the 
adjustment mechanism may explain the puzzling fact that the contribution of 
quintessence into $\Omega$ is close to 1. 

One can see from eq.~(\ref{wforphi}) that the lower limit for $w$ is $w>-1$
and this is quite generic for any normal matter. However in 
ref.~\cite{caldwell99} even a possibility of $w<-1$ was discussed with an
appropriate name "cosmic phantom". Such really striking equation of state
could be realized in models with higher rank tensor fields but it gives
rise to a very unusual cosmological singularity (see discussion in 
ref.~\cite{dolgov97b}).

\section{Neutrino.}

Neutrino as a possible candidate for dark matter has the following two 
advantages. First, it is the only one that is definitely known to exist.
Second, neutrino should have a non-zero mass. There are recent 
indications~\cite{sknu}  that at least one neutrino species has a mass 
about $0.07$ eV. However the second advantage
is simultaneously a disadvantage, because the neutrino mass is normally too
small for an appropriate description of the large scale structure of the 
universe. If cosmic background neutrinos of the $a$-th flavor have the standard 
cosmological abundance, $n_{\nu_a} = 3n_\gamma /11 \approx 112/ {\rm cm^3}$, 
then their mass is is restricted by the Gerstein-Zeldovich~\cite{gz} bound:
\be
\sum_a m_{\nu_a} < 94\, {\rm eV}\, \Omega h^2_{100}
\label{gz}
\ee
Such light neutrinos decoupled from cosmic plasma while they were relativistic
and they erased all structures by free streaming at the scales below
\be
M_{struc} \sim {m_{Pl}^3 \over m_\nu^2} 
\approx 10^{15} M_{\odot} \left(10\, {\rm eV} \over  m_\nu \right)^2
\label{mstrucnu}
\ee
This is typical example of a hot dark matter. (A more accurate estimate gives
somewhat smaller $M_{struc}$.)

On the other hand the Tremain-Gunn bound~\cite{tremain79} demands that 
neutrino mass is bounded from below:
\be
m_\nu > 50-100 \,\,{\rm eV}
\label{tg}
\ee
This bound is a striking example of quantum effects on galactic scale: Fermi
exclusion principle forbids too many neutrinos to accumulate in galactic
halo, hence to carry all observed mass they should be sufficiently heavy.

The mismatch between the bounds (\ref{gz}) and (\ref{tg}) does not allow 
the standard neutrinos to constitute all dark matter in the universe.
However, if neutrinos possess a new interaction
somewhat stronger than the usual electroweak one, their cosmological number
density would be smaller and the limit (\ref{gz}) would be less restrictive.
Another possibility is that there are the so called sterile neutrinos that
may be mirror or shadow neutrinos (see sec.~\ref{shadow}) with the mass in keV
range thus providing warm dark matter~\cite{berezhiani95}.

Some time ago a very heavy neutrino with the mass in GeV range was considered 
as a feasible candidate for cold dark matter. However the combined LEP
result~\cite{pdg} of precise measuring of $Z^0$ width
permits only $N_\nu = 2.993\pm 0.011$ for all neutral fermions with the 
normal weak coupling to $Z^0$ and mass below $m_Z/2 \approx 45$ GeV.  
So if heavy neutrinos, $\nu_h$, of the fourth generation exist their mass 
must be higher than 45 GeV. Most probably such particles should be unstable 
but if the corresponding leptonic charge is conserved or almost conserved
and the charged companion of the heavy neutrino is heavier than $\nu_h$
they would be stable or very long lived.

\begin{figure}[htb]
\begin{center}
  \leavevmode
  \hbox{
    \epsfysize=3.0in
    \epsffile{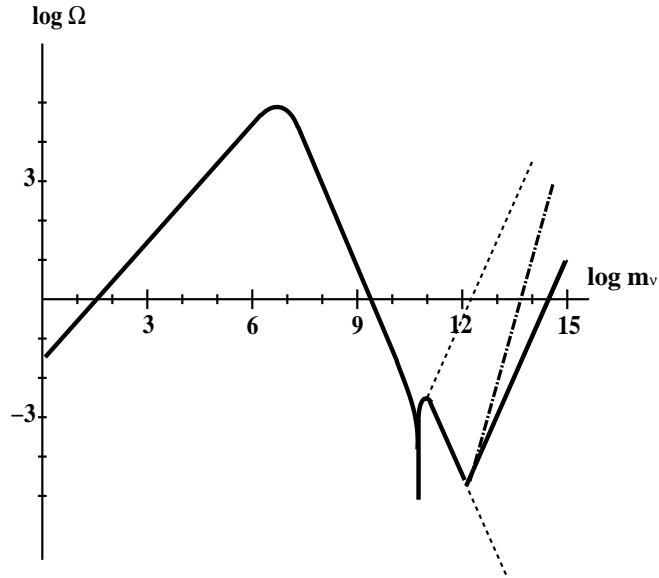}}
\end{center}
\caption{Contribution to cosmological parameter $\Omega$ from a heavy
stable neutrino as as function of its mass.
\label{figomeganu}}
\end{figure}

The contribution of $\nu_h$ into cosmological energy density is determined by
the cross-section of $\nu_h \bar \nu_h$-annihilation and has a rather
peculiar behavior as a function on the $\nu_h$ mass.  
The corresponding $\Omega$ is presented in fig.~\ref{figomeganu}.
In the region of very small masses the 
ratio of number densities $n_{\nuh}/n_\gamma$ does not depend upon the 
neutrino mass and $\rho_{\nuh}$ linearly rises with mass. This gives the
bound (\ref{gz}). For larger masses 
$\sigma_{ann} \sim \mnh^2$ and $\rho_{\nuh}\sim 1/\mnh^2$. This formally opens 
a window for $\mnh$ above 2.5 GeV~\cite{vdz,lee77}.
A very deep minimum in $\rho_{\nuh}$ near
$\mnh = m_Z /2$ is related to the resonance enhanced cross-section around
$Z$-pole. Above $Z$-pole the cross-section of $\bar \nuh \nuh$-annihilation  
into light fermions goes down with mass as $\alpha^2/\mnh^2$ (as in any normal
weakly coupled gauge theory). The corresponding rise in $\rho_{\nuh}$ is
shown by a dashed line. This would give the limit 
$\mnh < 3-5$ TeV~\cite{dz,olive82}.
However for $\mnh > m_W$ the contribution of the
channel $\bar \nuh \nuh \rar W^+W^-$ leads to the rise of the cross-section
with the increasing mass as 
$\sigma_{ann} \sim \alpha^2 \mnh^2 /m_W^4$~\cite{ekm}. This
would permit to keep $\rho_{\nuh}$ well below $\rho_c$ for all masses above
2.5 GeV. The behavior of $\rho_{\nuh}$ with this effect of rising cross-section
included, is shown by the solid line till $\mnh =1.5 $ TeV. Above that it
is continued as a dashed line. This rise with mass would break unitarity 
limit for partial wave amplitude when $\mnh$ reaches 1.5 TeV (or 3 TeV for 
Majorana neutrino)~\cite{gk,ek}.
If one takes the maximum value of the S-wave cross-section permitted by 
unitarity, which scales as $1/\mnh^2$, this would give rise
to $\rho_{\nuh} \sim \mnh^2$ and it crosses $\rho_c$ at $\mnh \approx 200$ TeV.
This behavior is continued by the solid line above 1.5 TeV. 
However for $\mnh \geq {\rm a\,\, few}\,\, TeV$ 
the Yukawa coupling of $\nuh$ to 
the Higgs field becomes strong and no reliable calculations of the annihilation
cross-section has been done in this limit. Presumably the cross-section is much 
smaller than perturbative result and the cosmological bound for $\mnh$ is close
to several TeV. This possible, though not certain, behavior is presented by 
the dashed-dotted line.

\section{Super-heavy relics.}

Super-heavy quasi-stable particles with the mass around $10^{13}$ GeV were
introduced recently in refs.~\cite{berezinsky97,kuzmin97,sarkar98} 
to avoid the GKZ-cutoff~\cite{gkz}
for ultra-high energy cosmic rays. These particles could have
produced at the end of inflation by coherent oscillations of the
inflaton field (for possible mechanisms of production see e.g.
ref.~\cite{chang98,kuzmin98}). 
Some cosmological and astrophysical
constraints on superheavy quasistable relics were discussed  
earlier in refs.~\cite{ellis92}. Such particles 
may have an interesting impact on 
structure formation and are discussed in more details in this conference
by H. Ziaeepour. However their meta-stability is rather mysterious. As was
argued many years ago by Zel'dovich~\cite{zel76}, even if baryonic charge
is microscopically conserved, proton may decay through formation and 
subsequent evaporation of a virtual black hole.  
In accordance with his estimate the proton should decay with the
life-time:  
\be
\tau_p \approx {1\over m_p}\left({ m_{Pl} \over m_p }\right)^4
\approx 10^{45} {\rm years}
\label{taup}
\ee
This estimate can be obtained as follows. The cross-section of the
gravitational capture of a particle by the black hole with mass $M$ is
equal to its Schwarzschild radius squared,
\be
\sigma_{grav} \approx r_g^2 = {M^2 \over m_{pl}^4}
\label{sigma}
\ee
where $m_{Pl} = 1.2\times 10^{19}$~GeV is the Planck mass.
For the virtual black hole state, which is formed in the process of the 
gravitational decay of a particle with mass $m$, the mass of black hole
is around the initial particle mass, $M\sim m$. Assuming that all other
dimensional parameters are also close to $m$ we obtain the result 
(\ref{taup}).  

We can obtain another (and different) estimate for the proton life-time 
using the following arguments. The amplitude of the collapse of a
particle $x$ with mass $m_x$ into 
black hole with the same mass is proportional to the overlap integral:
\be
A_{coll} \sim \int d^3 r \psi_x \Psi_{BH}
\label{overlap}
\ee
where $\psi_x$ and $\Psi_{BH}$ are the wave functions of the particle and 
black hole. The particle wave function is localized on its Compton wave length,
$l_C = 1/m_x$, while the black hole wave function is localized at 
$r_g = m_x/m_{Pl}^2$. Evaluating this integral and assuming again that all
other dimensional parameters are close to $m_x$ we obtain
\be
\tau_x \sim {1\over m_x} \left( { m_{Pl} \over m_x }\right)^n
= 10^{-24+19n} \left({{\rm GeV}\over m_x} \right)^{n+1} {\rm sec}
\label{taux}
\ee
where the power $n$ is equal to 6, in contrast to $n=4$ in eq.~(\ref{taup}).

Later on this conjecture was supported by the arguments that quantum
gravity effects should break all global symmetries~\cite{glob}, in
particular due to formation of baby universes~\cite{baby}. Effective
Lagrangian which describes these phenomena contains different terms
with different powers of Planck mass, $m_{Pl}^{4-d}$, where $d$ is called
the dimension of the corresponding operator. In the examples considered above
$d$ was equal to 6 and 7. The very dangerous terms are those with $d=5$.
They would lead to the proton decay with life-time $\tau_p \sim 10^{13}$~sec,
which is well below existing limits. This makes one to believe that the
operator with $d=5$ do not appear in effective Lagrangian. Note, that the
simple estimates presented above give $d>5$. If the particle decay is
generated by the operator with dimension $d$ then its life-time is given by
the expression~(\ref{taux}) with $n=2(d-4)$. Thus if we demand that the
particle $x$ lives longer that the universe age, $t_U \approx 10^{18}$~sec,
then its mass should be bounded from above:
\be
m_x < 10^{{\left(19n-42\right) / \left( n+1\right) } } \,\,{\rm GeV}
\label{mx}
\ee   
If the Zeldovich estimate~\cite{zel76} is correct then $m_x < 10^7$~GeV. 
If we use the estimate of the present paper which gives $n=6$, then 
$m_x < 10^{10.3}$~GeV. The condition that these particles are heavier than
$10^{13}$~GeV, so that their decays explain the origin of 
ultra-energetic cosmic rays, demands a rather high value $n> 9$. The dimension
of the corresponding operators should be bigger than 8.5. 
 
Of course the arguments presented above are not rigorous but still the
gravitational decay mechanism looks very plausible.   
This mechanism is quite generic and does not depend upon the 
particle properties but only on their masses. This is related to the 
universality of gravitational interactions. Of course the presented 
estimates are rather naive and the unknown non-perturbative dynamics of 
quantum gravity may significantly change these results. It is possible in
particular that the formation of a virtual black hole proceeds as some
kind of tunneling process. In this case the decay probability might be
suppressed as $\exp ( -c\, m_{Pl} /m_x)$ (where $c$ is a constant)
and the discussed here mechanism would be ineffective. 

A possible way to avoid the gravitational decay is to assume that the 
particle in question is the lightest in the family of particles
possessing a conserved charge, which is associated with a local (gauge)
symmetry (similar to electromagnetic $U(1)$). However it would imply that 
this particle is absolutely stable. To avoid that one would have to assume 
that the corresponding gauge symmetry is slightly broken in such a way
that the gauge boson(s) acquires a tiny but non-zero mass. It is well known
that black holes may have only hairs which are related to the long range 
forces which in turn are associated with zero mass of the particles which
transmit interactions. For example Coulomb field of electrically charged
black hole is maintained outside the gravitational radius only because
photon is strictly massless. In the case that photon has a non-zero mass, a
black hole would not have electric hairs even if electric charge is strictly
conserved. A limiting transition from the case of strictly massless photon to
that with a small mass is achieved by a long time of disappearance of the 
hairs. This time should be inversely proportional to the mass. So in principle
there may exist very heavy and very long lived particles if they possess a
conserved charge but the corresponding gauge symmetry is a little broken 
so that the gauge boson acquire a tiny mass. The charge may remain strictly 
conserved but the particle would be unstable in the same way as proton 
becomes unstable due to the collapse into black hole, despite conservation
of baryonic charge in particle interactions without gravity. A possible way
to realize such a model is to assume a nonminimal and gauge non-invariant
coupling of gauge bosons to gravity, for example in the form $A_\mu^2 R$
or $A_\mu A_\nu R^{\mu\nu}$, where $R$ is the curvature scalar and 
$R^{\mu\nu}$ is the Ricci tensor.

Barring this a highly speculative possibility we have to conclude that either
the explanation of the highest energy cosmic rays by decays of ultra-heavy
long-lived particles is impossible, because such particles should undergo
fast ($\tau_x < t_U$) decay or that the  gravitational  
breaking of global symmetries is not as strong as we assumed above.

\section{Lightest supersymmetric particle (LSP).}

Low energy supersymmetry has at least two attractive features for a solution
of dark matter problem. First, the theory predicts an existence of new 
stable particles that could constitute cosmological dark matter. Second,
with a natural scale of supersymmetry breaking around 1 TeV, the theory
predicts that LSP would give $\Omega_{LSP} \approx 1$ without any
fine tuning. The third feature, that makes this hypothesis especially 
attractive for experimentalists, is that for a large range of parameters of 
supersymmetric models these new stable particles are within the reach of 
of sensitivity of different existing and planned methods of their
search. This subject was recently reviewed in great detail in 
ref.~\cite{bottino99,masiero99}, so I will be very brief here.

There are several possible candidates for the role of the dominating 
supersymmetric matter in the universe: neutralino (a mixture of gauginos,
$\tilde \gamma + \tilde Z$, and higgsinos, $\tilde h_1 + \tilde h_2$);
sneutrino (a heavy supersymmetric partner of neutrino);
gravitino (the supersymmetric partner of graviton, with spin 3/2);
axino (the partner of axion), messenger fields related to a hidden sector 
of the theory, ... . Such  particles (at least some of them) can be searched 
for directly by a registration in low background detectors (Ge, NaI, Xe,...)
through the reaction: $ N + {\rm Nuclei} \rar {\rm recoil}$. There are
also indirect methods based on search for the products of their annihilation
in the Earth or in the Sun, producing high energy muons. At the present day 
only upper limits on the annihilation cross-section are established, though 
there are indications on annual modulation effect~\cite{dama} that may be
a signature of dark matter. 

A very interesting feature of neutralino annihilation in the galactic halo 
is a production of antimatter: not only anti-protons~\cite{antip}
but also a noticeable fraction of anti-deuterium may be created.
According to calculations of ref.~\cite{antid} the flux of $\bar D$ at 
low energy, below 1 GeV, would be much larger than the flux of the
secondary $\bar D$, produced by the normal cosmic ray collisions.  
The AMS mission could either register anti-deuterium from neutralino
annihilation or exclude a significant fraction in the parameter space of 
the low energy SUSY models. There are also promising ways to register
neutralino annihilation through observation of energetic positrons or
gamma rays (see ref.~\cite{bottino99} for the details).

A low energy supersymmetric extension of the minimal standard model is 
very natural from particle physics point of view.
It supplies possibly the best candidate for the dark 
matter particles. In most versions of the model these particles would form 
weakly interacting cold dark matter, though in some cases
warm dark matter is also possible.  
There is a very high experimental activity in search of supersymmetric
particles and hopefully at the beginning of the next millennium they will 
be discovered or, if the nature is not favorable, a large part of the 
parameter space will be excluded but the mystery of dark matter will still  
remain.

\section{Mirror/shadow world \label{shadow}}

The idea that our world is doubled and there exists a similar or exactly the
same world coupled to ours only by gravity,  was suggested long  
ago~\cite{lee56} in connection with conservation of parity, P, or combined
parity, CP. Subsequently it was developed and elaborated in several 
papers~\cite{mirror}. Its popularity greatly increased after it was found
that superstring theories have $G\times G$ internal symmetry group and the 
two identical worlds, corresponding to two groups, communicate only through
gravity~\cite{green89}. The considered models, however, were not confined
to this simplest option. In addition to gravity a new super-weak (but 
stronger than gravity) interaction was introduced  between our and 
mirror particles. Moreover, a different patterns of symmetry breaking in these
two worlds were considered, so that physics in our and in the mirror, or 
in this case better to say in the shadow world, became quite different.  

At first sight the existence of a whole new world with the same or similar
particle content would strongly distort successful predictions of the
standard big bang nucleosynthesis (BBN) theory. The latter permits 
not more than
one additional light fermionic species in the cosmological plasma at 
$T\sim 1$ MeV (see e.g.~\cite{olive99}). The completely symmetric mirror 
world would give slightly more than 7. However, as was argued in 
ref.~\cite{kolb85} the temperature of the mirror matter after inflation 
could be smaller than the temperature of the usual matter and thus the
energy density of mirror matter during nucleosynthesis could be safely 
suppressed. Concrete mechanisms that could create a colder
mirror world if the symmetry between the worlds was broken,
were considered e.g. in refs~\cite{berezhiani95,berezhiani96}. Another
possible way to escape a conflict with BBN by the generation of lepton
asymmetry through neutrino oscillations was discussed in ref.~\cite{foot97}.

A new burst of interest to mirror/shadow matter arose after MACHO 
collaboration announced that the mass of the micro-lenses, they observed, 
is close to the solar mass (see sec.~\ref{baryon}). A natural idea that
these objects may be built from mirror matter, immediately attracted a strong
attention~\cite{berezhiani95,berezhiani96,blinnikov98,foot99,mohapatra99}.
In the case of exact symmetry between the worlds the properties of the 
stellar objects would be the same but the process of structure formation 
could be quite different by the following two reasons. First, since the 
the mirror matter is colder than the usual one, the mirror hydrogen 
recombination would be considerably earlier and the structures might start
forming earlier too. Second, baryon asymmetry in the mirror world might be
different from ours and it would have an important impact on primordial 
chemical content of the universe and galactic and 
stellar formation~\cite{comber}.
The cosmological mass fraction of mirror baryons is unknown but most
probably they do not constitute all dark matter in the universe. There is
one peculiar feature of this matter that it is strongly interacting and
can easily loose energy through emission of mirror photons. Structure
formation with this kind of dark matter would be very much different from 
the normal scenario with dissipationless cold dark matter. The cooling
mechanisms, that are very essential for structure formation, could be either
stronger or weaker. In particular, in the world with a very large fraction 
of mirror $^4He$ molecular cooling would be considerably less efficient.  

There would be even more difference between cosmology and astrophysics of our
and mirror world if the mirror symmetry is 
broken~\cite{berezhiani95,berezhiani96}. There could be the case that 
there are no stable nuclei in the mirror world and thus there could not exist 
mirror stars with thermonuclear active core. If the mirror electrons are 
heavier than the usual ones, the mirror hydrogen binding energy would be 
larger and this would be another reason for earlier recombination. To study
the history of stellar formation and evolution in such distorted
world would be a very interesting exercise that could reveal essential 
features of the underlying physics. Except for a different astrophysics
and new stellar size invisible bodies, mirror world could provide sterile
neutrinos that might explain the observed neutrino anomalies though the 
oscillations between our neutrinos and sterile ones. In particular,
among these sterile neutrinos there could be rather heavy ones with the
mass in keV range that might be excellent candidates for warm dark matter.

\section{Miscellanea \label{misc}}

Because of lack of space and time I could not discuss many other interesting
forms of dark matter. One of the favorites, axion, is discussed at this  
conference by Yu. Gnedin. Topological and non-topological solitons may be 
also quite interesting options. Though the measurements of the angular
fluctuations of CMB seemingly exclude cosmic strings as dominant part of
cosmological dark matter, they still may give some contribution to the 
total mass of the universe. Non-topological solitons, $Q$-balls, recently
attracted a renewed attention~\cite{kusenko,enqvist}. Primordial black
holes with the log-normal mass spectrum~\cite{dolgov93} still remain an
interesting possibility. There are some even more exotic candidates that
are discussed in the literature; among them are such objects as 
superstrings giving super-heavy dark matter~\cite{dvali99}, 
domain walls with "anti-gravitating" equation of state, 
$p=-(2/3 )\rho$~\cite{battye99}, or even liquid or solid dark 
matter~\cite{eichler96}.

Unstable dark matter remains attractive, and though it was proposed at
the beginning of 80~\cite{unstab84}, the main burst of activity
was in the 90th~\cite{unstab90}. The basic idea of introducing unstable but
long-lived particles into consideration was to increase the horizon length
at the time of equality between matter and radiation and to increase by
that the power at large scales. Recently this idea was revived in another 
attempt to save a model of structure formation with pure cold dark 
matter~\cite{masiero99a}. The model looks quite natural from particle physics
point of view if there exists a light scalar boson, familon or majoron
so that a heavier neutrino, that may violate Gerstein-Zeldovich bound, could
decay into this boson and lighter neutrino. It is also possible  
that a massive scalar boson decays into two light neutrinos. A very interesting 
scenario in the former case is that the scalar bosons are massive and 
their spectrum is two component: energetic bosons coming from the decay and 
non-relativistic ones formed during phase transition similar to axions. In 
this case the same particle may form both cold and hot (or warm) dark matter.
A slightly different mechanism 
was proposed in ref.~\cite{brustein99} in the frameworks of string
cosmology. It was argued there that 
weakly interacting non-thermal relics may be produced in the course of dilaton 
driven inflation with the double peak spectrum that could simultaneously give 
cold and hot dark matter.

A very interesting form of dark matter is a self-interacting one. One
possible example of the latter is given by mirror or shadow world discussed
above. A few more models of self-interacting dark matter with particles 
belonging to our world were considered in the literature; they were
either light bosons~\cite{self}, e.g. majorons or familons, or neutrinos 
with an anomalous self-interaction~\cite{atrio97}. Observational evidence 
in favor of self-interacting dark matter was recently analyzed in
ref.~\cite{spergel99}

\section{Conclusion \label{conclusion}}

As we have seen, a set of independent arguments unambiguously proves that the
main part of matter in the universe is not visible and, moreover, this 
invisible matter is not the matter that consists of known elementary 
particles as e.g. protons or neutrons, or neutrinos. Existence of this
unknown form of matter is a strong evidence in favor of new physics beyond
the minimal standard $SU(3)\times SU(2)\times U(1)$-model (MSM). Possibly 
a low energy supersymmetric extension of MSM solves the mystery of dark matter
with lightest supersymmetric particle (LSP) that quite probably could be 
stable. However astronomical data indicate that one form of dark matter is 
not enough and except for cold dark matter, that 
might be provided by LSP, there is a very
strong quest for hot and/or warm dark matter. Moreover detailed description
of rotation curves at small distances indicates that dark matter may be
dissipative. Quite possibly there is 
one more ingredient of dark matter, related to vacuum energy, that makes the
situation even more mysterious.

Even if there is only one form of dark matter, the cosmic 
conspiracy, namely the close values of $\Omega_{baryon}$ and $\Omega_{DM}$,
is quite puzzling. It demands quite a strong fine-tuning in the fundamental
particle theory and at the present day no reasonable understanding of the
phenomenon exists. The problem of cosmic conspiracy becomes tremendously
deeper if there are several $(>2)$ forms of invisible matter with the
similar contributions to $\Omega$. 

An answer to an often asked question, what is the best bet for the dark matter 
particles, reflects not so much our knowledge of the subject but a personal
attitude of the respondent. Seemingly most votes would be given to LSP and
possibly the next one is the axion. An advantage of these two is that both 
were not invented {\it ad hoc} but were 
predicted by particle theory independently of cosmology. By similar 
arguments mirror or shadow matter is also in a good shape. However other
candidates based on more complicated models may have better chances just
because their properties are chosen in accordance with cosmological demands.

10 years ago in one of "Rencontre de Moriond" meeting P. Peebles in his
summary talk arranged a public opinion pool, how many dark matter candidates
would survive to the end of the century. The stakes were up to double digit 
numbers. I have to admit that I voted for one dark matter candidate,
the only real one that "would be surely known". It was extremely over-optimistic 
point of view and today we have even more possible candidates than 10 years
ago (neither old ones is removed from the list and quite a few new ones came
into being) and still do not know what is/are the correct one(s).

{\bf Acknowledgments}
The work of A.D. was supported by Danmarks Grundforskningsfond through its
funding of the Theoretical Astrophysical Center.


\end{document}